\documentclass[namedreferences]{solarphysics}

\usepackage[hyperref,optionalrh,showbiblabels]{spr-sola-addons} 
\usepackage{graphicx}        
\usepackage{amssymb}        
\usepackage{color}           
\usepackage{breakurl}        

\usepackage{multirow} 
\usepackage{array, booktabs, makecell} 

\chardef\us=`\_

\begin{document}

\begin{article}
\begin{opening}

\title{Characteristics of the Solar Coronal Line Profiles from Fabry-Perot Interferometric Observations}

\author[addressref={aff1},corref,email={maya.prabhakar@iiap.res.in}]{\inits{Maya}\fnm{Maya}~\lnm{Prabhakar}\orcid{https://orcid.org/0000-0003-2141-4184}}
\author[addressref=aff1,email={kpr@iiap.res.in}]{\inits{K.P.}\fnm{K.P.}~\lnm{Raju}\orcid{https://orcid.org/0000-0002-6648-4170}}
\author[addressref=aff2,email={chandra@prl.res.in}]{\inits{T.}\fnm{T.}~\lnm{Chandrasekhar}}

\address[id=aff1]{Indian Institute of Astrophysics, Bangalore}
\address[id=aff2]{Physical Research Laboratory, Ahmedabad}

\runningauthor{M. Prabhakar, K.P. Raju and T. Chandrasekhar}
\runningtitle{Characteristics of the Solar Coronal Line Profiles}

\begin{abstract}

This article reports the analysis of a set of Fabry-Perot interferograms that were studied to probe the physical parameters of the inner solar corona.
The observations were carried out in the coronal green line, Fe {\tiny XIV} 5302.86 \AA, during the total solar eclipse of 21 June 2001 which occurred in Lusaka, Zambia.
The study was performed in the radial range of 1.1\textemdash1.5 \(\textup{R}_\odot\) and examines the Doppler velocity, halfwidth, centroid and asymmetry and their correlations with each other at various points in the corona.
It is found that 59 \% of the line profiles are blueshifted, 34 \% of them are single components and only 7 \% of them are redshifted.
The variations of halfwidth and Doppler velocity with respect to coronal height have a large scatter and shows no significant changes. 
It has been found that the variations of halfwidth with Doppler velocity or centroid follow a parabolic trend with a weak correlation, whereas the relation between halfwidth and the asymmetry is inconclusive.
These results may provide more insight to the coronal dynamics and help in understanding the coronal physical problems.

\end{abstract}
\end{opening}

\section{Introduction}
     \label{S-Introduction} 
The transition region and the solar corona are widely explored for their several intriguing features. 
The dynamics of the flows and mass motions, existence of multiple component line profiles, line shifts and their dominance, non-thermal velocities and intensity variations at different regions are a few topics that are widely discussed.
Most of these studies are aimed to find if they can provide any insight to the coronal heating and acceleration of the solar winds.
The solar corona has been an interesting region having a very low density of the order of $10^{8-9}$ cm$\rm^{-3}$
and a temperature of the order of 2 MK.
Many theories like nanoflare heating, type-II spicules, magnetic reconnection, Alfven waves are proposed to explain this high temperature \citep{2017PJAB...93...87S}.
Study continues in search of non-thermal sources of energy that could cause coronal heating and acceleration of the solar winds.  
The study of the solar coronal line profiles is one such method where we can gain information about the physical parameters of the solar corona.

Asymmetries have been generally reported in the spectral lines from the transition region and the solar corona.
Mass motions in the corona has been reported by \cite{1969SoPh....9..116D,1975SoPh...45..157D,1988JApA....9..125D} and \cite{1991SoPh..131...25C}.
The presence of the multiple components with excess blueshifts was reported by \cite{1993MNRAS.263..789R}.
\cite{1995ApJ...455L..81B} mention the increasing fraction of redshifts in the transition region. 
\cite{1997SoPh..175..349B} and \cite{1999ESASP.446..645T} reported that the transition region comprises of redshifts.
\cite{1998ApJS..114..151C} and \cite{1999ApJ...522.1148P} also found excess blueshifts through Solar Ultraviolet Measurements of Emitted Radiation (SUMER)  observations. 
\cite{1999SoPh..185..311R} showed the existence of multi-components in the line profiles.
\cite{2011A&A...534A..90D} also found excess blueshifted profiles in the corona.
\cite{2006ApJ...647.1452P} explain these emissions in terms of nanoflare heating and \cite{2009ApJ...701L...1D} related them to type-II spicules. 
\cite{2012ApJ...760L...5B} explain these asymmetries to be signatures of chromospheric jets which provide mass and energy to the corona.

Further, there has been difference in the results regarding the variation of the non-thermal velocity and linewidth with height. 
The linewidth of the red line is reported to increase with height above the limb by  \cite{2006JApA...27..115S}.
\cite{2013SoPh..282..427P} present a negative gradient of the green line width with height. 
\cite{2004A&A...427..725R} report about the anisotropy of the velocity distribution in this region.
The nature of the variation of the line profile parameters is still unclear, and in this paper, we aim to study it in more detail.

In this context, we have focused on the solar coronal physical parameters like Doppler velocity, halfwidth, centroid and asymmetry and their correlations at various points away from the limb by analyzing a set of Fabry-Perot interferograms.
We have primarily focused on the variation of Doppler velocity and centroid of the line widths of the line profiles.
This work is a continuation of our earlier work \citep{2013ASInC..10..137P} which mainly focused on the shifts and asymmetries of the line profiles.
Section~\ref{S-Analysis and data reduction} gives details of the instrument used, analysis, and data reduction methods involved in our study.
In Section~\ref{S-Results and discussion}, we mention the results found in our study and discuss them in comparison to the earlier results. Section~\ref{S-Conclusion} gives the conclusion of our study.

\section{Analysis and Data Reduction}  
\label{S-Analysis and data reduction}

Our study involves the analysis of 14 Fabry-Perot interferograms.
These were taken during the total solar eclipse of 21 June 2001 that occurred in Lusaka, Zambia.
It had a magnitude of 1.0495 and lasted for 3 min 37 sec when the Sun was at 31$^{\circ}$ elevation above the northwest horizon. 
From the data, we have obtained intensity, Doppler velocity, halfwidth, centroid and asymmetry at different points in the corona. 
The centroid is defined as the wavelength point that divides the area of the line profile into two \citep{2011ApJ...736..164R}.
To calculate the asymmetry, strips of equal width (0.5 {\AA}) at equal distances (0.1 {\AA}) from the peak wavelength are considered in the red (\textit{R}) and blue (\textit{B}) regions.
The asymmetry is then calculated using the relation (\textit{R-B})/(\textit{R+B}). 
The observation is done in the coronal green line { Fe {\tiny XIV} 5302.86 \AA}, which is chosen as its formation temperature (2 MK) is close to the average inner coronal temperature, which means that the results yield more accurate information about the corona.

\begin{table}
\caption{ Instrumental features of the Fabry-Perot interferometer used for our study.
}
\label{Table1}
\begin{tabular}{lll}     
  \hline                   
 Features & Values \\
  \hline
  
Free spectral range & 4.75 {\AA} \\
Instrumental width & 0.2 {\AA}  \\
Spectral resolution & 26,000 \\
CCD configuration & 1024 x 1024\\
Pixel size & 24 $\mu$m \\
Pixel resolution & 3.4 arcsec \\

   \hline
\end{tabular}
\end{table}

The Fabry-Perot interferometer used is the one used in  \cite{1984ApOpt..23..508C}.
However, during the 21 June 2001 eclipse, a CCD was used as the detector.
The interferometer has a free spectral range of 4.75 {\AA}, an instrumental width of 0.2 {\AA}, a spectral resolution of 26,000 for the coronal green line and a pixel resolution of 3.4 arcsec. 
The instrumental details are given in Table~\ref{Table1}. 
The line profiles have been extracted in the angular range of 240$^{\circ}$. 
The original plan of the experiment was to get a time-sequence of Fabry-Perot interferograms of the solar corona  during the total solar eclipse. 
The aim was to study the temporal changes of the physical properties of the solar corona.
But due to a minor tracking error that happened during the observation, it was later noticed that the interferograms are not co-spatial. 
Hence, we could not study the temporal variations.
However, this enabled us to study the corona at more spatial locations than was originally planned. 

In a Fabry-Perot interferogram, the region between two adjacent fringe minima constitutes a line profile.
Hence the spatial resolution in an imaging Fabry-Perot interferometer is generally low. 
It is about 0.2 \(\textup{R}_\odot\) in the present observation.
The tracking error in the observation will shift an interferogram with respect to the previous one, leading to the formation of a new fringe pattern in the gap.
This will give line profiles from new locations albeit with some boxcar averaging of the nearby spatial points.
Hence the angular coverage remains the same (240$^{\circ}$) as in a single interferogram, but we get more number of line profiles from the intermediate regions.
Part of the data obtained during this particular eclipse was used by \cite{2011ApJ...736..164R} and \cite{2013ASInC..10..137P}.

\begin{figure}    
   \centerline{\hspace*{0.03\textwidth}
               \includegraphics[width=0.515\textwidth,clip=]{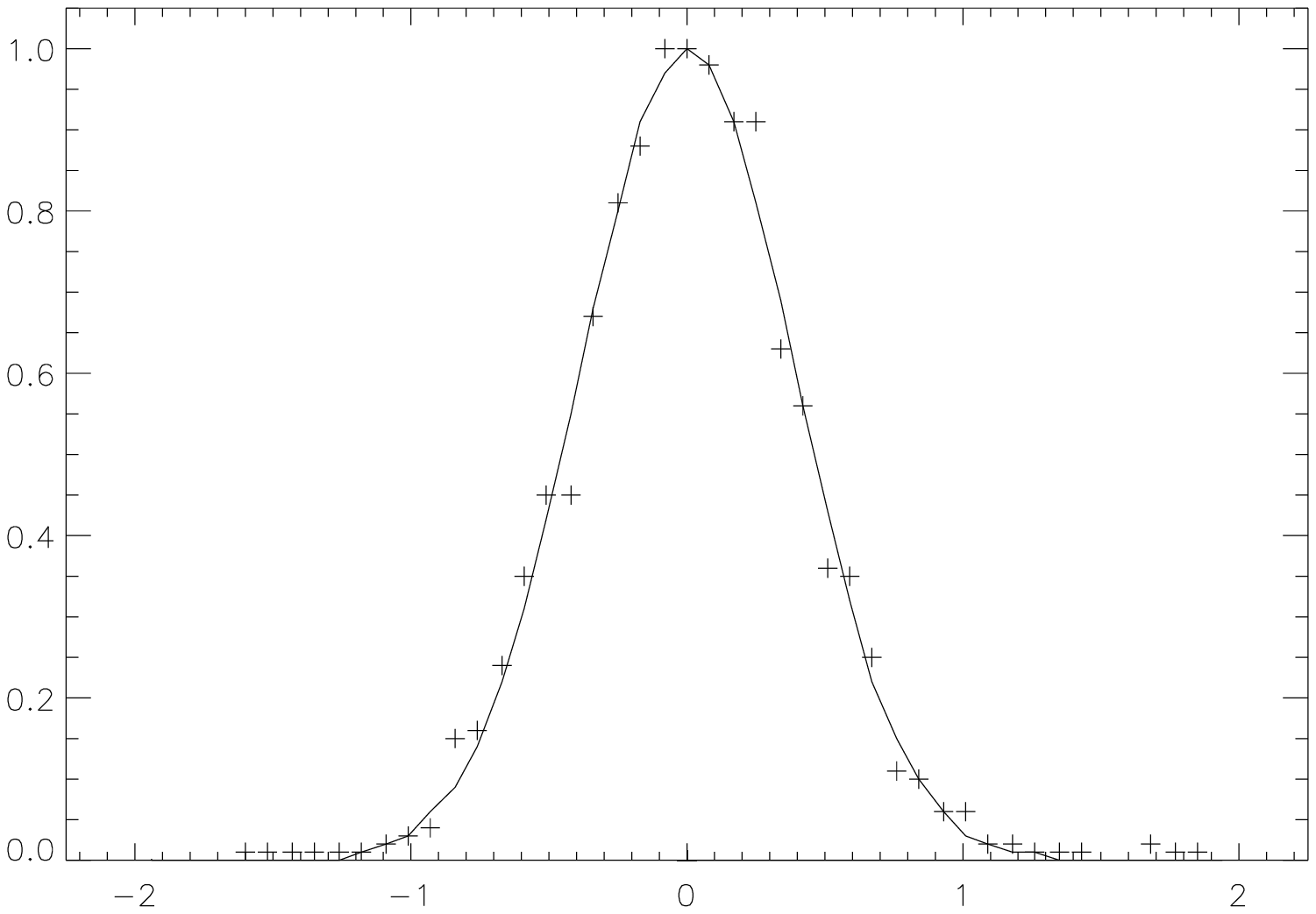}
               \hspace*{-0.05\textwidth}
               \includegraphics[width=0.515\textwidth,clip=]{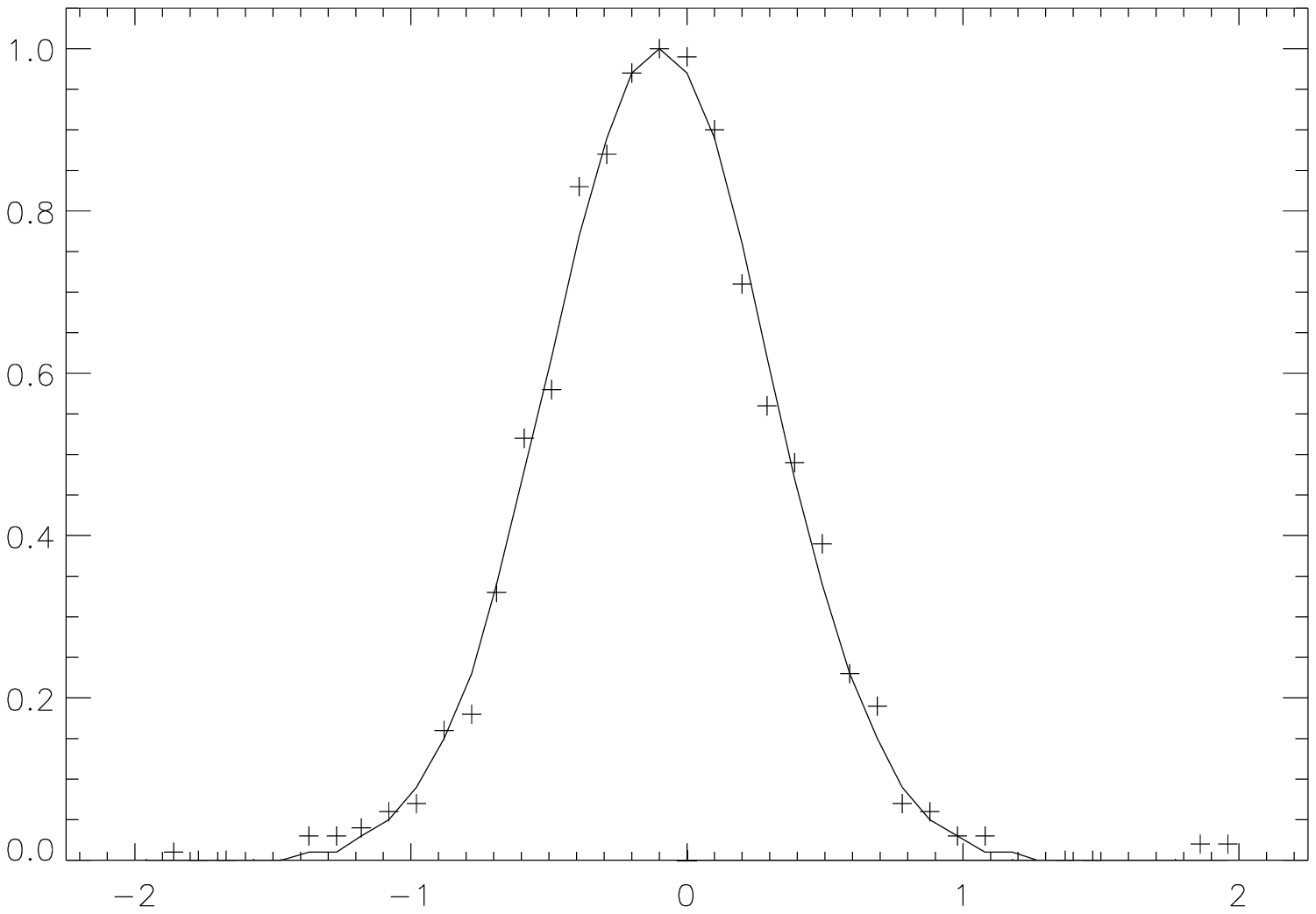}
               \hspace*{-0.01\textwidth}
              }
     \vspace{-0.4\textwidth}   
     \centerline{\Large \bf     
      \hspace{0.0 \textwidth}  \color{white}{(a)}
      \hspace{0.415\textwidth}  \color{white}{(b)}
         \hfill}
     \vspace{0.31\textwidth}    
   \centerline{\hspace*{0.03\textwidth}
               \includegraphics[width=0.515\textwidth,clip=]{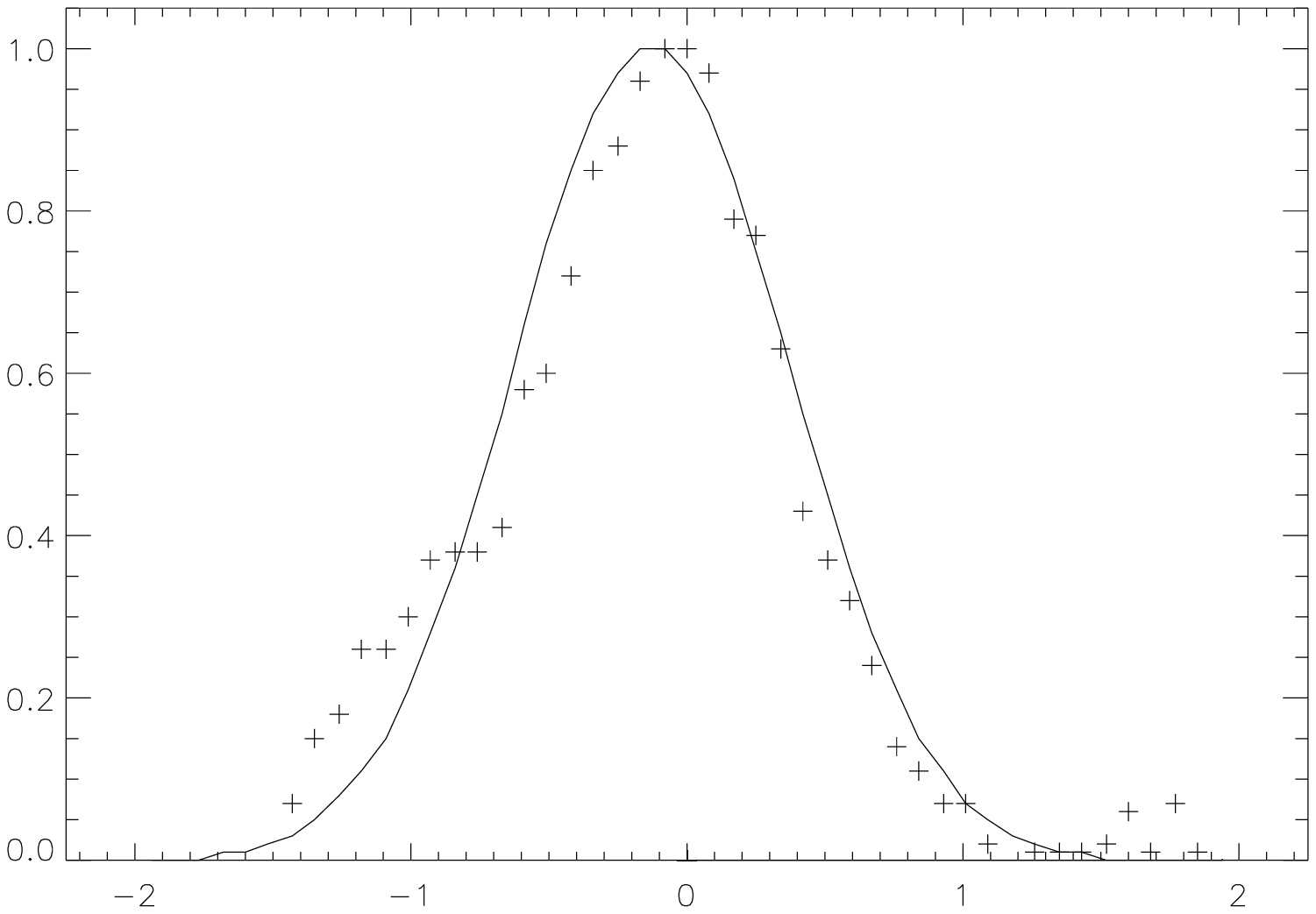}
               \hspace*{-0.05\textwidth}
               \includegraphics[width=0.515\textwidth,clip=]{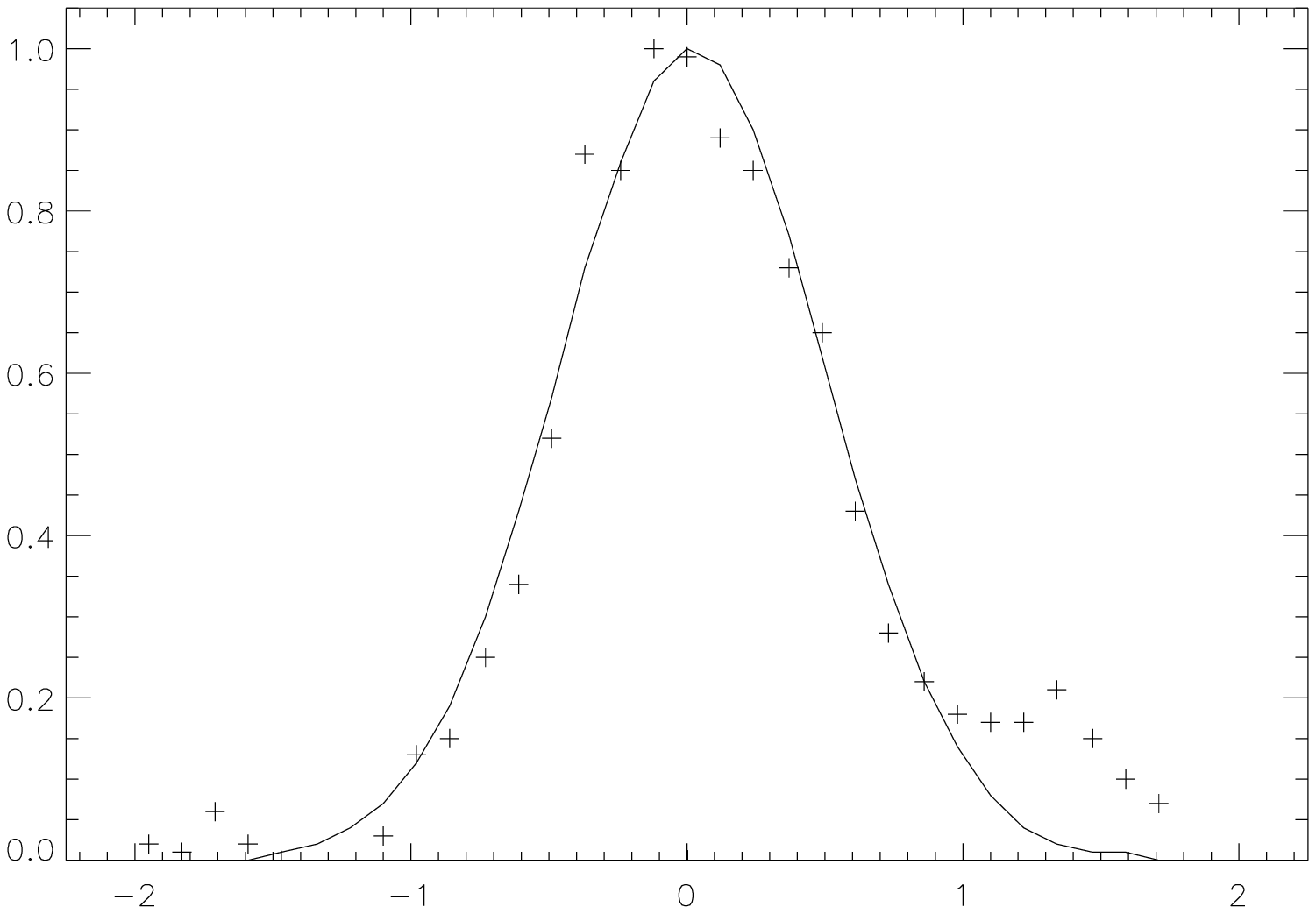}
               \hspace*{-0.01\textwidth}
              }
     \vspace{-0.35\textwidth}   
     \centerline{\Large \bf     
      \hspace{0.0 \textwidth} \color{white}{(c)}
      \hspace{0.415\textwidth}  \color{white}{(d)}
         \hfill}
     \vspace{0.31\textwidth}    
              
\caption{Example of line profiles fitted with Gaussian curves. The upper two line profiles represent single components profiles and the lower two represent blueshifted (left) and redshifted (right) line profiles.
        }
   \label{F-Figure1}
   \end{figure}

\begin{figure}    
   \centerline{\includegraphics[width=0.5\textwidth,clip=]{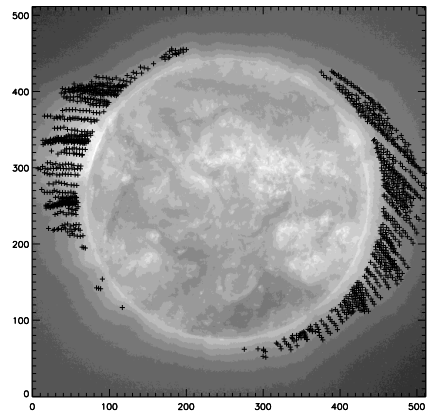}
              }
              \caption{EIT image of the Sun on 21 June 2001 at time 13:48:13  showing the spatial locations of all the line profiles analyzed. 
                      }
   \label{F-figure2}
   \end{figure}

The analysis is done by first locating the fringe center position in the interferograms. 
Then the radial scans from the fringe center were made to obtain the line profiles.
The wavelength is calibrated as in \cite{1993MNRAS.263..789R}.
In order to reduce the noise, a 2x2 pixel averaging was done at the outset. It was also noticed that the line profiles close to the limb were slightly contaminated by the scattered light in some interferograms. 
Therefore we have considered only those line profiles beyond 1.1 \(\textup{R}_\odot\).
Single Gaussian curves are fitted to all the line profiles and the parameters are obtained.
We have selected only those line profiles whose signal-to-noise ratio is $\geq$ 15 in order to get good Gaussian fits. 
A set of line profiles fitted with Gaussian curves is shown in Figure~\ref{F-Figure1}.
The estimated errors in the fitting are about 5\% in intensity, 2 km s$^{-1}$ 
in velocity and 0.03 {\AA}  in width.
Interactive Data Language (IDL) is used for the computation purpose. 
All the spatial locations of the line profiles obtained are plotted on an Extreme ultraviolet Imaging Telescope (EIT) image of the Sun obtained during the same time as that of the eclipse, which can be seen in Figure~\ref{F-figure2}.

\section{Results and Discussion}
\label{S-Results and discussion}

We obtained 1295 line profiles in all by removing the noisy profiles.
We found that 59 \% of them are blueshifted, 7 \% are redshifted and 34 \% are single components. 
Single components are those which fall in the asymmetry range of -0.08\textemdash0.08 \AA\ corresponding to the Doppler velocity range of -2\textemdash2 km s$^{-1}$
It is interesting to see that the percentages of the single components, blueshifts and redshifts are different from what is observed in  \cite{2011ApJ...736..164R} and \cite{2013ASInC..10..137P}.
Note that the above works had considered close to 300 line profiles.
In this work, the improved statistics give a better reliability (a factor of about 2) to our results as compared to the above mentioned works.  

This work is primarily focused on the variations of the Doppler velocity, halfwidth, centroid, asymmetry, and their interrelationships. 
The gross properties of a large number of line profiles fitted with a single Gaussian were obtained. 
By examining the interrelationships between the quantities,  we expect to get information on the nature of the multicomponents without actually going for the complex multiple Gaussian fitting. 
In \cite{2011ApJ...736..164R}, line profiles were obtained from a single interferogram and the nature of the secondary component, obtained by subtracting the red wing from blue, was studied in detail. 
In \cite{2013ASInC..10..137P}, the study is primarily on the asymmetry of the multiple components.

The number of line profiles observed in different asymmetry ranges is given in Table~\ref{Table2}.
It can be seen that the line profiles with the greatest asymmetry are found in the blue wing (96 line profiles). 
Thus, our results agree well with the works by \cite{1993MNRAS.263..789R}, \cite{1999SoPh..185..311R}, \cite{2011A&A...534A..90D}, \cite{1998ApJS..114..151C} and \cite{1999ApJ...522.1148P} regarding the presence of the multicomponents and dominance of the blueshifts, but the percentage is found to be on the higher side. 
The redshifts are seen to be very less with just 7 \% contribution.

\begin{table}
\caption{ Table showing number of line profiles observed in different ranges of asymmetry.
}
\label{Table2}
\begin{tabular}{lccc}     
 \hline

 Type & Asymmetry range ({\AA}) & Number of profiles & Total\\

\hline
    
 Single &-0.08 to 0.08 & 439 & 34\% \\
 \\
 Blueshifts & -0.08 to -0.2 & 428 \\
               & -0.2 to -0.3 & 237 \\
               & -0.3 to -0.6 & 96 & 59\% \\
   \\
 Redshifts & 0.08 to 0.2 & 76 \\
              & 0.2 to 0.3 & 19 & 7\% \\
  \hline
\end{tabular}
\end{table}

\begin{figure}    
   \centerline{\hspace*{0.015\textwidth}
               \includegraphics[width=0.515\textwidth,clip=]{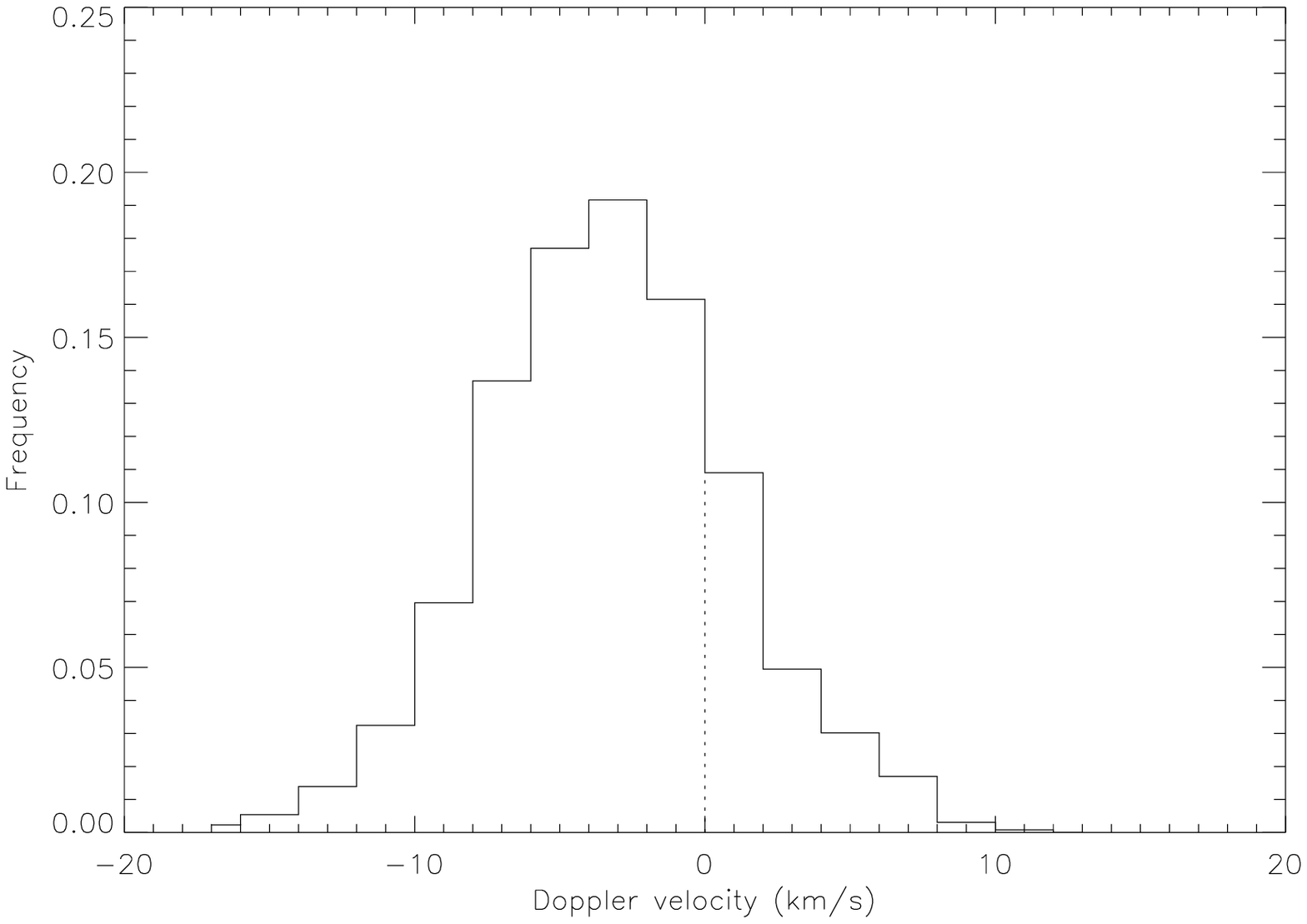}
               \hspace*{-0.03\textwidth}
               \includegraphics[width=0.515\textwidth,clip=]{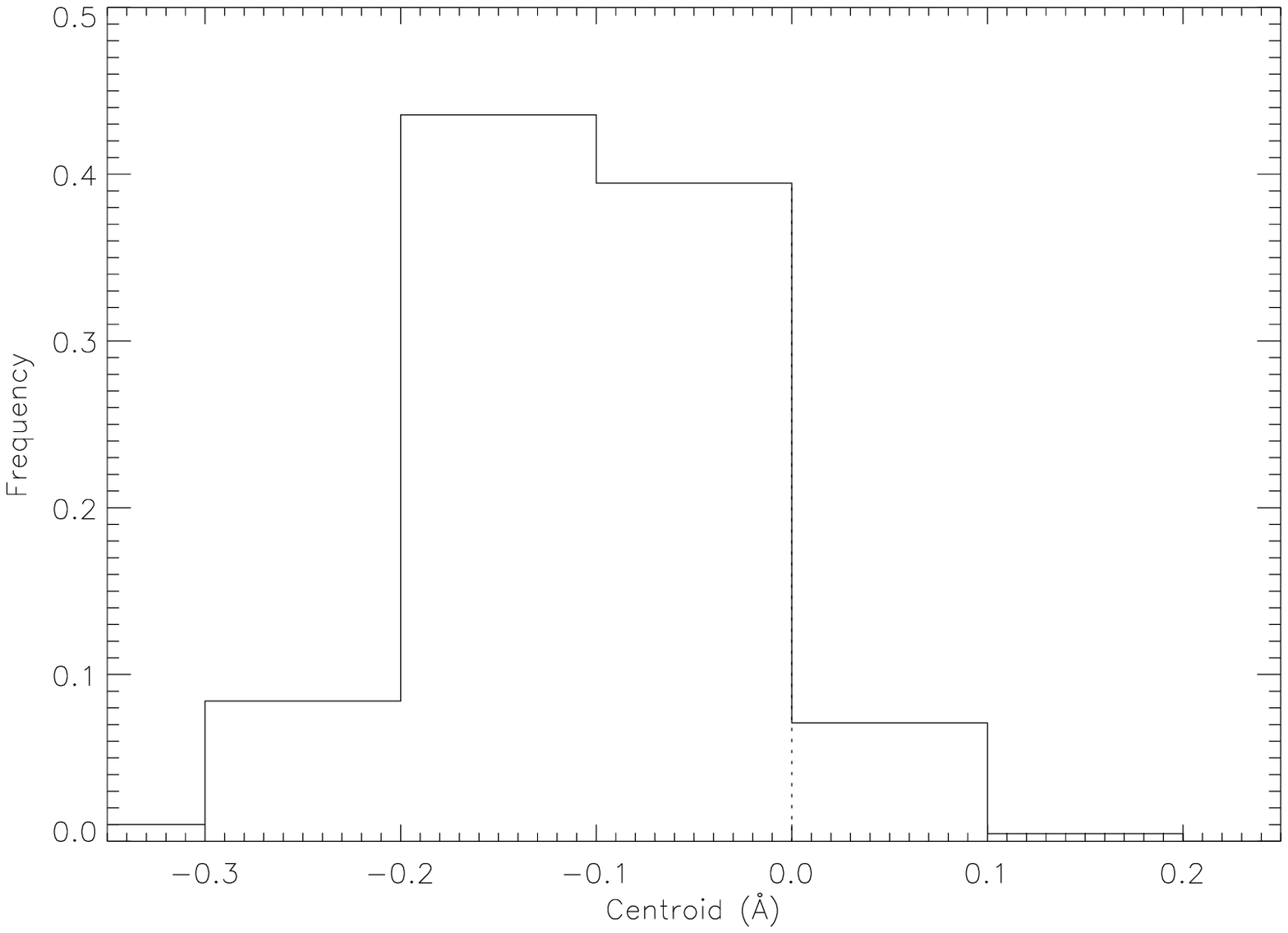}
              }
     \vspace{-0.34\textwidth}   
     \centerline{\Large \bf     
      \hspace{0.068 \textwidth}  \color{black}{(a)}
      \hspace{0.415\textwidth}  \color{black}{(b)}
         \hfill}
     \vspace{0.31\textwidth}    
               \centerline{\includegraphics[width=0.5\textwidth,clip=]{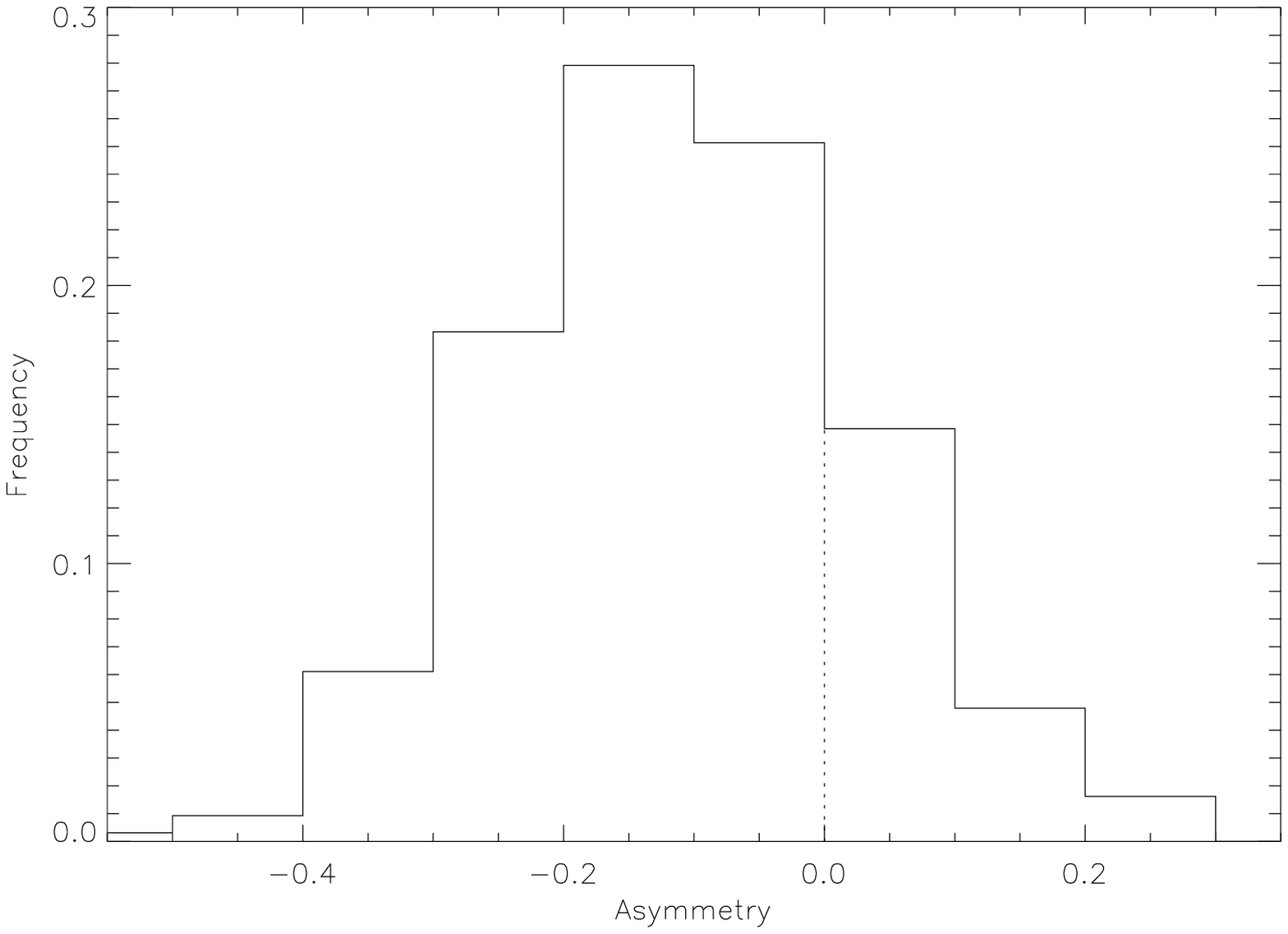}
              }
        \vspace{-0.33\textwidth}   
     \centerline{\Large \bf     
      \hspace{0.310 \textwidth} \color{black}{(c)}
      \hspace{0.415\textwidth}  \color{white}{(d)}
         \hfill}
     \vspace{0.31\textwidth}    
 
\caption{Normalized histograms of (a) Doppler velocity, (b) Centroid, and (c) Asymmetry.
        }
           \label{F-figure3}
   \end{figure}

\begin{figure}    
   \centerline{\includegraphics[width=0.5\textwidth,clip=]{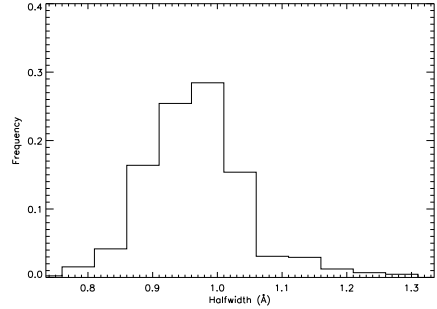}
              }
              \caption{Normalized histogram of halfwidth of the line profiles.
                      }
   \label{F-figure4}
   \end{figure}

\begin{figure}    
   \centerline{\hspace*{0.015\textwidth}
               \includegraphics[width=0.515\textwidth,clip=]{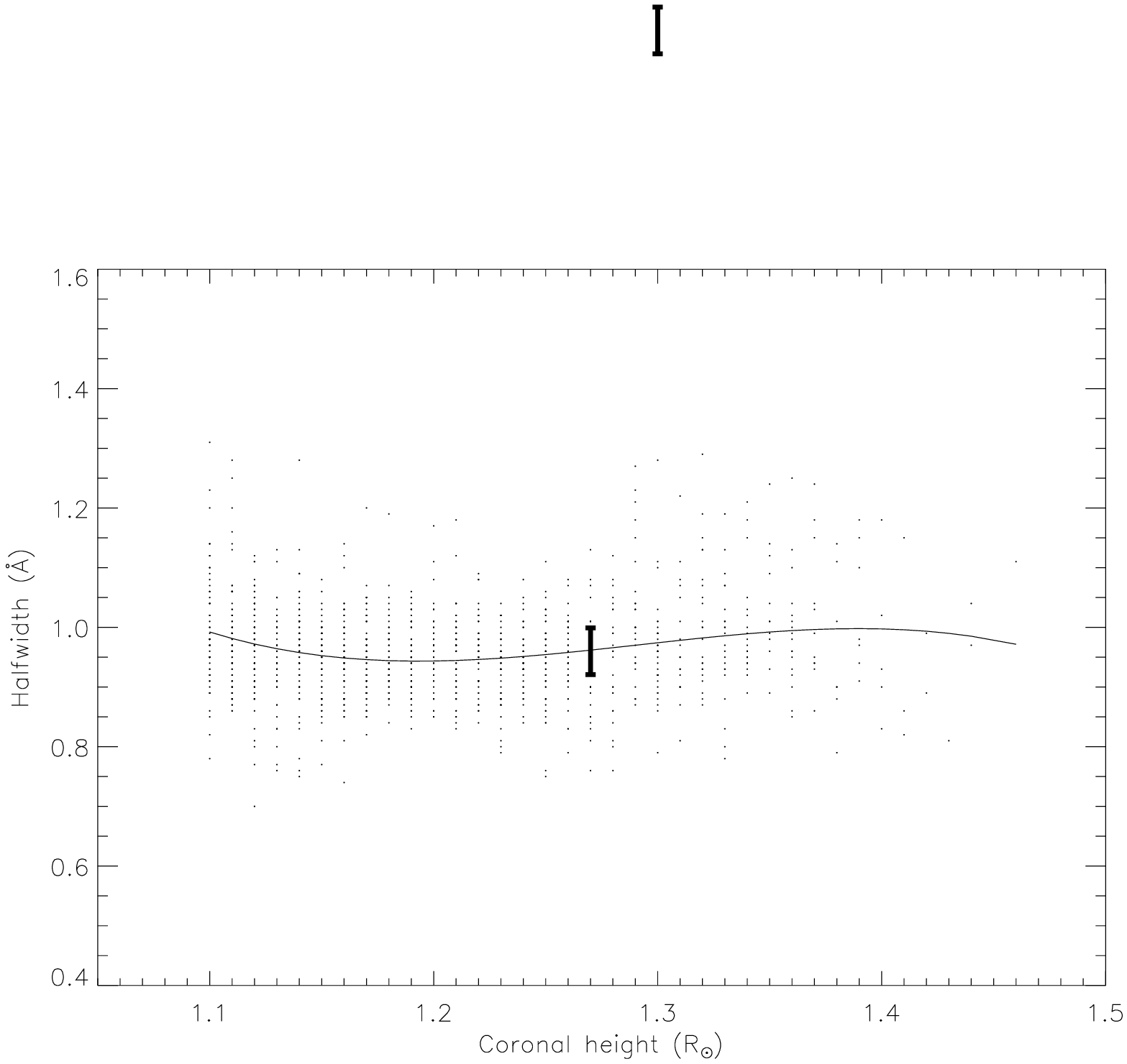}
               \hspace*{-0.03\textwidth}
               \includegraphics[width=0.515\textwidth,clip=]{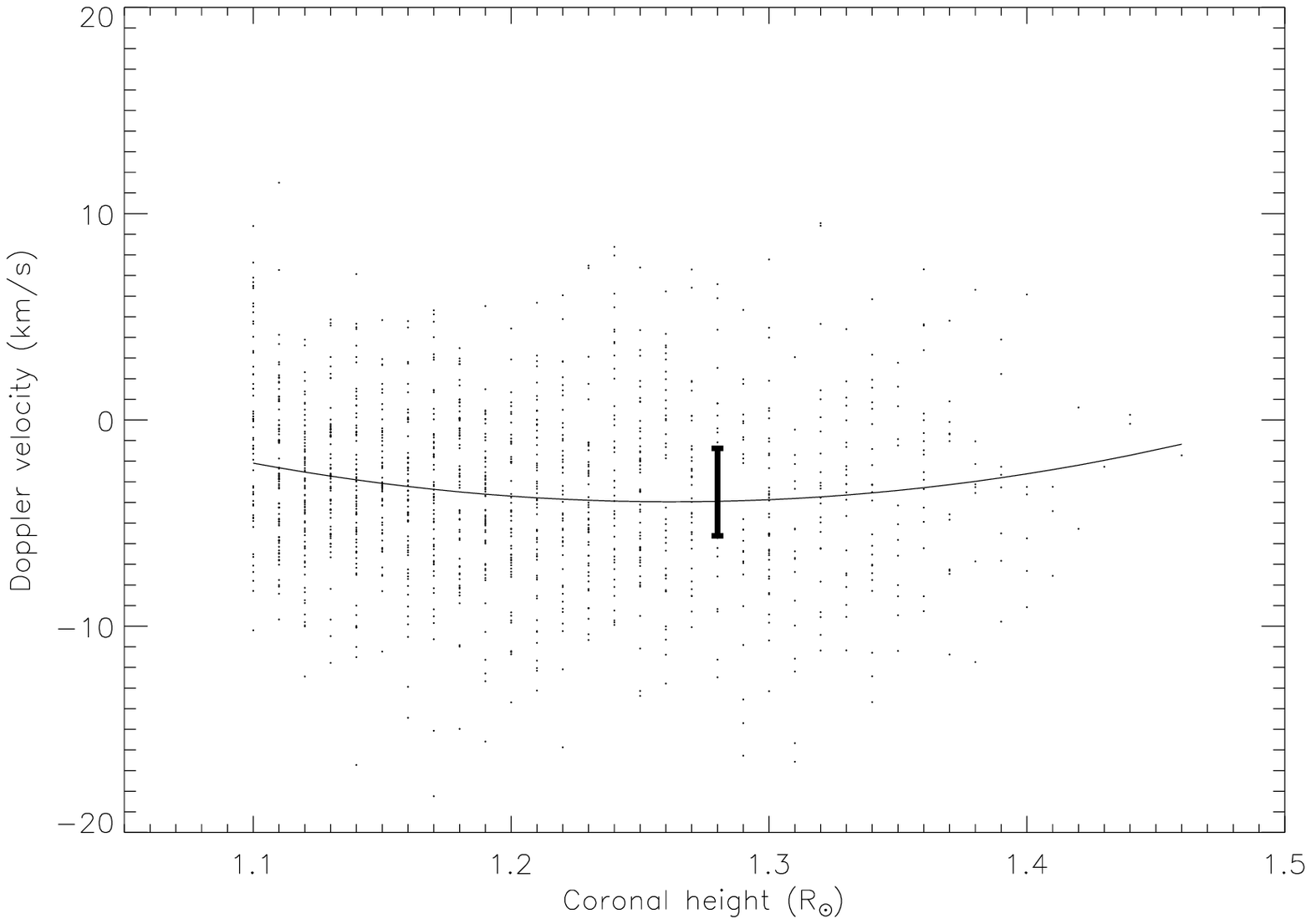}
              }
     \vspace{-0.34\textwidth}   
     \centerline{\Large \bf     
      \hspace{0.06 \textwidth}  \color{black}{(a)}
      \hspace{0.415\textwidth}  \color{black}{(b)}
         \hfill}
     \vspace{0.31\textwidth}    

\caption{(a) Variation of the halfwidth and (b) Doppler velocity with height above the limb. The solid line shows the best fit polynomial. The error bar indicates the standard error in the fittings.
        }
           \label{F-figure5}
   \end{figure}

The normalized histograms of Doppler velocity, centroid and asymmetry of the line profiles can be seen in Figure~\ref{F-figure3}. 
It can be seen that all of them show the domination of the blueshifts which does not agree with  \cite{2012ApJ...749...60M}, which report a weak emission component in the blue wing that contributes a very less percentage of the emission line profiles. 
The multi-components with blueshifts are found to have a maximum Doppler velocity of -18 km s$^{-1}$
and those with redshifts are found to have a maximum Doppler velocity of 11 km s$^{-1}$ 
It is to be noted that these values are from the composite line profiles and when the Gaussian decomposition is performed, velocities of the secondary components are much larger \citep{2011ApJ...736..164R}.

The normalized histogram of the halfwidth can be seen in Figure~\ref{F-figure4}. It can be seen that the halfwidth fall in the range 0.7-1.3 {\AA} and peak at 0.96 {\AA}. 
This would correspond to a temperature of 3.5 MK. Considering the non-thermal broadening that could add to the line width, and the line formation temperature as 2 MK, we find that the non-thermal velocity is close to 22 km/s \citep{2011ApJ...736..164R} from the following relation:

\begin{eqnarray}   
\frac{2kT_\textup{0}}{M} &=& \frac{2kT_\textup{D}}{M} \ + \ v_\textup{t}^2      \nonumber
                     \\      
                     \nonumber 
\end{eqnarray}

\begin{table}
\caption{ Table showing values of \textit{R}, \textit{R$^{2}$} and standard errors for the plots in Figures in 5 and 6 for linear, quadratic and cubic fittings.
}
\label{Table3}
\tabcolsep=0.15cm
\begin{tabular}{ cccccccccc }     
 \hline

Figure & \multicolumn{3}{c}{Linear } & \multicolumn{3}{c}{Quadratic} & \multicolumn{3}{c}{Cubic}\\

  & \textit{R} & \textit{R$^{2}$} & {\thead{Standard\\ error \\} } & \textit{R} & \textit{R$^{2}$} & {\thead{Standard\\ error \\ }} & \textit{R} & \textit{R$^{2}$} & {\thead{Standard\\ error \\ }} \\
 \cline{1-10}\\
 
 Figure 5a & 0.05 & 0.0025 & 0.07 & 0.18 & 0.031 & 0.078 & 0.21 & 0.044 & 0.078 \\
      \\
 Figure 5b & -0.08 & 0.0068 & 4.26 & 0.14 & 0.018 & 4.23 & 0.14 & 0.021 & 4.23 \\
      \\ 
 Figure 6a & 0.03 & 0.0008 & 0.079 & 0.23 & 0.051 & 0.077 & 0.22 & 0.048 & 0.077 \\ 
      \\
 Figure 6b & 0.02 & 0.0004 & 0.079 & 0.21 & 0.045 & 0.078 & 0.21 & 0.045 & 0.078 \\     
      \\
 Figure 6c & 0.12 & 0.015 & 0.079 & 0.12 & 0.014 & 0.079 & 0.12 & 0.015 & 0.079 \\
 
  \hline
\end{tabular}
\end{table}

where k is the Boltzmann constant, T$_\textup{0}$\ is the observed line width temperature, M is the mass of the emitting ion, T$_\textup{D}$\ is the line width Doppler temperature and v$_\textup{t}$\ is the non-thermal velocity characterizing microturbulence.

Further, we studied the variation of the halfwidth and Doppler velocity of the line profiles with respect to their heights from the solar center.
In Figure~\ref{F-figure5}a, we have plotted halfwidth against the coronal height.
We have tried linear, quadratic and cubic polynomial fits and a  detailed statistical analysis was performed.
The details are given in Table~\ref{Table3}.
The different columns in the table give the values of correlation coefficient (\textit{R}),  coefficient of determination (\textit{R$^{2}$}), and the standard error in the fittings.
It can be seen from the table that the best fit is obtained from the cubic polynomial fit where the value of \textit{R}=0.21 and \textit{R$^{2}$}=4.4\%. The best fit polynomial and the standard error are plotted in the figure. It can be seen that the total variation is comparable to the standard error in the fitting and hence the trend is insignificant.

This result is unlike what is observed in \cite{2008A&A...480..509M}, which states that the width remains almost constant or increases up to a height of 1.3 \(\textup{R}_\odot\). 
It is also different from what is seen in \cite{2013SoPh..282..427P} and \cite{2016SoPh..291.2281B}, which report a decrease in the line widths with height of the line profiles.

Figure~\ref{F-figure5}b shows the variation of the Doppler velocity of the line profiles with coronal height.
It can be seen from Table~\ref{Table3} that the coefficients \textit{R} and \textit{R$^{2}$} do not show much improvement from quadratic to cubic fits.
So, a quadratic fit is considered for this plot.
Here the total variation is less than the standard error and hence insignificant.

\begin{figure}    
   \centerline{\hspace*{0.015\textwidth}
               \includegraphics[width=0.515\textwidth,clip=]{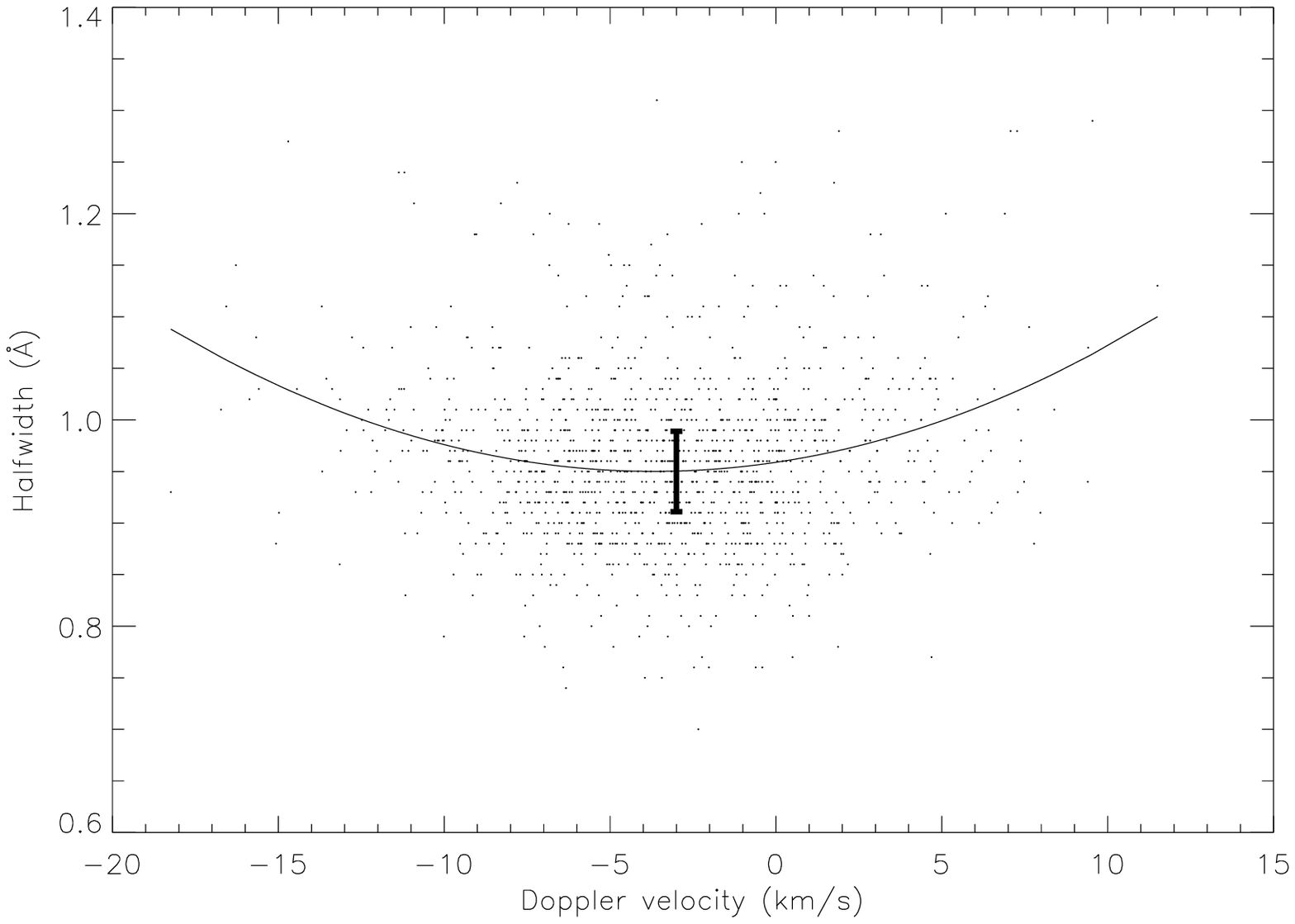}
               \hspace*{-0.03\textwidth}
               \includegraphics[width=0.515\textwidth,clip=]{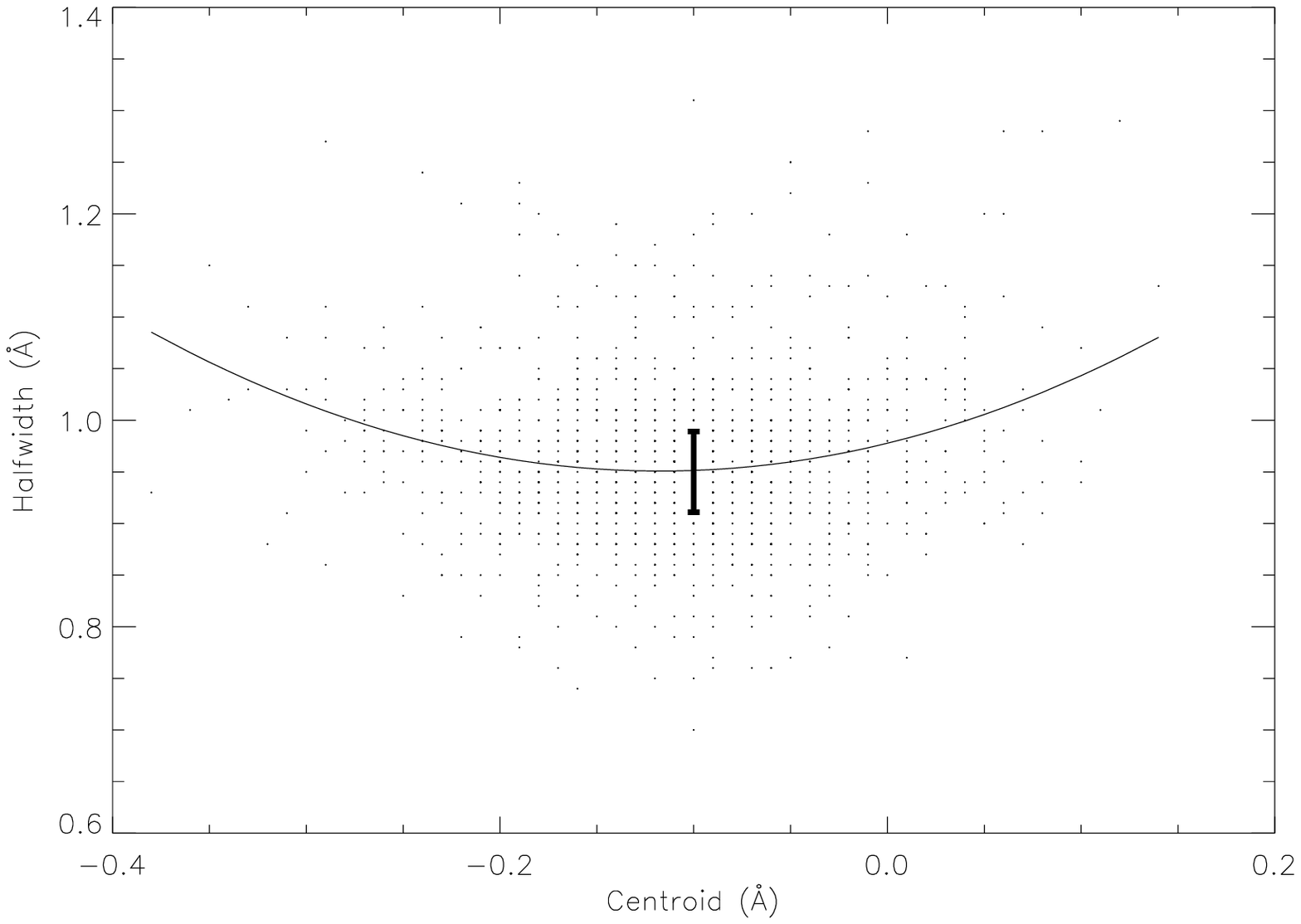}
              }
     \vspace{-0.34\textwidth}   
     \centerline{\Large \bf     
      \hspace{0.068 \textwidth}  \color{black}{(a)}
      \hspace{0.415\textwidth}  \color{black}{(b)}
         \hfill}
     \vspace{0.03\textwidth}    
         
            \vspace{0.31\textwidth}    
               \centerline{\includegraphics[width=0.5\textwidth,clip=]{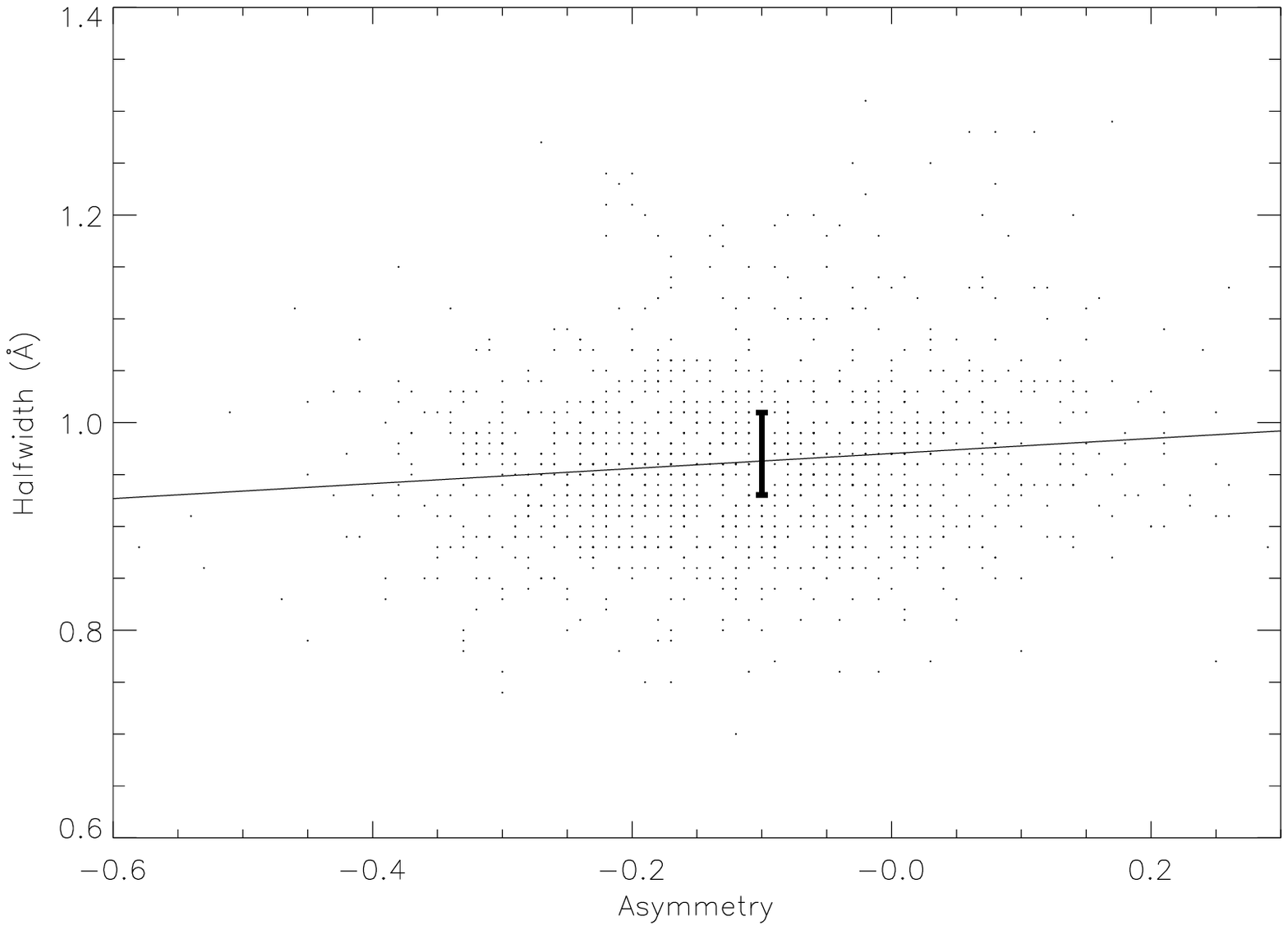}
              }
        \vspace{-0.33\textwidth}   
     \centerline{\Large \bf     
      \hspace{0.31 \textwidth} \color{black}{(c)}
      \hspace{0.415\textwidth}  \color{white}{(d)}
         \hfill}
     \vspace{0.31\textwidth}    
 
\caption{(a) Variation of the Doppler velocity (b) Centroid, and (c) Asymmetry of the line profiles with halfwidth. The solid line shows the best fit polynomial. The error bar indicates the standard error in the fittings.
        }
           \label{F-figure6}

            \end{figure}

We further studied the variation of halfwidth of the line profiles with Doppler velocity and centroid which are shown in Figure~\ref{F-figure6}a and b. 
From Table~\ref{Table3}, it is evident that the the quadratic fit is better than the linear fit where the coefficients show a marked improvement.
There is no marked improvement in the cubic fit.
The total variation is almost twice of the standard error and hence these fits are retained. The correlation coefficients are 0.23 and 0.21 respectively which means a mild correlation. 
The \textit{R$^{2}$} values which gives the percentage of explained variation are about 5\%.
This imply that the relationship between halfwidth and Doppler velocity or centroid is weak.
A variation like this is observed in \cite{2010ApJ...715.1012B} for the active region AR 10978.
\cite{2011ApJ...736..164R} examined the relationship between halfwidth and Doppler velocity from two cases with limited number of samples and a positive angle coverage of 20$^{\circ}$. They report a weak correlation in one case and a lack of correlation in the other. Therefore, it may be concluded that the weak correlation found between the halfwidth and Doppler velocity or centroid could be real. \cite{2010A&A...521A..51P} reports that the positive correlation between these quantities suggest a flow associated with a heating process.

Variation of the halfwidth with respect to the asymmetry of the line profiles is given in Figure~\ref{F-figure6}c. Here the quadratic or cubic fit do not improve the linear fit.
The coefficients \textit{R}, \textit{R$^{2}$} and standard error remains almost constant in all the three cases. The linear trend is comparable to the standard error and is not significant. 
This is found very different when compared to the variation of Doppler velocity and centroid with the halfwidth.

\section{Conclusion}
\label{S-Conclusion}

We studied various characteristics of the inner solar corona by analyzing the line profiles obtained from the analysis of the Fabry-Perot interferograms.
We found that a majority of these line profiles comprise of multicomponents, with a higher contribution from the blueshifts (59 \%).
This was followed by the single component profiles, which constituted 34 \%.
The redshifts were found very less with just 7 \%.
This result, though found to be in agreement with the related works reported earlier, shows very high percentage of the blueshifts. 
Line profiles showing the greatest asymmetry were found to be blueshifted and they constituted close to 7.5\% of the observed line profiles. 
The multiple component profiles with blueshifts were found to have a maximum Doppler velocity of -18 km s$^{-1}$
and the redshifted multicomponents were found to have a maximum Doppler velocity of 11 km s$^{-1}$ 
These values specify the Doppler velocity of the composite profile and the actual values of the Doppler velocity of the components could be much higher. 
The excess blueshifts could be related to the type-II spicules or the nascent solar wind flow.
Further, the variation of halfwidth and Doppler velocity with respect to the coronal height are found to be insignificant. The variation of halfwidth with respect to the Doppler velocity or centroid  show a parabolic trend with a weak correlation. The positive correlation is important as it suggests a flow associated with a heating process. The trend between halfwidth and asymmetry is inconclusive which is not understood. 
These results need to be examined further for understanding the problems in coronal physics.

\begin{acks}
The authors would like to thank the reviewer for his valuable comments and suggestions. Maya Prabhakar would like to thank the Women Scientist Scheme - A (WOS-A) of the Department of Science \& Technology for providing financial support to carry out this project. This  work  is also  funded  by  the  Department  of  Science  \& Technology and the Department of Space, Government of India.\\

\end{acks}

{\textbf{Disclosure of Potential Conflicts of Interest:} The authors declare that they have no conflicts of interest.}

\bibliographystyle{spr-mp-sola}
\bibliography{references}

\begin{thebibliography}{27}
\ifx\bisbn     \undefined \def\bisbn  #1{ISBN #1}\fi
\ifx\binits    \undefined \def\binits#1{#1}\fi
\ifx\bauthor   \undefined \def\bauthor#1{#1}\fi
\ifx\batitle   \undefined \def\batitle#1{#1}\fi
\ifx\bjtitle   \undefined \def\bjtitle#1{\textit{#1}}\fi
\ifx\bvolume   \undefined \def\bvolume#1{\textbf{#1}}\fi
\ifx\byear     \undefined \def\byear#1{#1}\fi
\ifx\bissue    \undefined \def\bissue#1{#1}\fi
\ifx\bfpage    \undefined \def\bfpage#1{#1}\fi
\ifx\blpage    \undefined \def\blpage #1{#1}\fi
\ifx\burl      \undefined \def\burl#1{\textsf{#1}}\fi
\ifx\href      \undefined \def\href#1#2{\textsf{#2}}\fi
\ifx\betal     \undefined \def\betal{\textit{et al.}}\fi
\ifx\bctitle   \undefined \def\bctitle#1{#1}\fi
\ifx\beditor   \undefined \def\beditor#1{#1}\fi
\ifx\bbtitle   \undefined \def\bbtitle#1{\textit{#1}}\fi
\ifx\bedition  \undefined \def\bedition#1{#1}\fi
\ifx\bseriesno \undefined \def\bseriesno#1{\textbf{#1}}\fi
\ifx\blocation \undefined \def\blocation#1{#1}\fi
\ifx\bsertitle \undefined \def\bsertitle#1{\textit{#1}}\fi
\ifx\bsnm      \undefined \def\bsnm#1{#1}\fi
\ifx\bsuffix   \undefined \def\bsuffix#1{#1}\fi
\ifx\bparticle \undefined \def\bparticle#1{#1}\fi
\ifx\barticle  \undefined \def\barticle#1{}\fi
\ifx\binstitute  \undefined \def\binstitute#1{#1}\fi
\ifx\bpublisher  \undefined \def\bpublisher#1{#1}\fi
\ifx\doiurl    \undefined
  \def\doiurl#1{\href{http://dx.doi.org/#1}{\textsf{DOI}}}\fi
\ifx\arxivurl  \undefined
  \def\arxivurl#1{\href{http://arxiv.org/abs/#1}{\textsf{arXiv}}}\fi
\ifx\adsurl    \undefined
  \def\adsurl#1{\href{http://adsabs.harvard.edu/abs/#1}{\textsf{ADS}}}\fi
\ifx\botherref \undefined \def\botherref#1{}\fi
\ifx\url       \undefined \def\url#1{\textsf{#1}}\fi
\ifx\bchapter  \undefined \def\bchapter#1{}\fi
\ifx\bbook     \undefined \def\bbook#1{}\fi
\ifx\bcomment  \undefined \def\bcomment#1{#1}\fi
\ifx\oauthor   \undefined \def\oauthor#1{#1}\fi
\ifx\citeauthoryear \undefined\def \citeauthoryear#1{#1}\fi
\ifx\endbibitem\undefined \def\endbibitem{}\fi
\ifx\bconflocation  \undefined \def\bconflocation#1{#1} \fi

\bibitem[\protect\citeauthoryear{{Beck}
  \textit{et~al.}}{2016}]{2016SoPh..291.2281B}
\begin{barticle}
\bauthor{\bsnm{{Beck}}, \binits{C.}},
\bauthor{\bsnm{{Rezaei}}, \binits{R.}},
\bauthor{\bsnm{{Puschmann}}, \binits{K.G.}},
\bauthor{\bsnm{{Fabbian}}, \binits{D.}}:
\byear{2016},
\batitle{{Spectroscopy at the Solar Limb: II. Are Spicules Heated to Coronal
  Temperatures?}}
\bjtitle{\solphys}
\bvolume{291},
\bfpage{2281}.
\doiurl{10.1007/s11207-016-0964-4}.
\adsurl{2016SoPh..291.2281B}.
\end{barticle}
\endbibitem

\bibitem[\protect\citeauthoryear{{Brekke}, {Hassler}, and
  {Wilhelm}}{1997}]{1997SoPh..175..349B}
\begin{barticle}
\bauthor{\bsnm{{Brekke}}, \binits{P.}},
\bauthor{\bsnm{{Hassler}}, \binits{D.M.}},
\bauthor{\bsnm{{Wilhelm}}, \binits{K.}}:
\byear{1997},
\batitle{{Doppler Shifts in the Quiet-Sun Transition Region and Corona Observed
  with SUMER on SOHO}}.
\bjtitle{\solphys}
\bvolume{175},
\bfpage{349}.
\doiurl{10.1023/A:1004985226553}.
\adsurl{1997SoPh..175..349B}.
\end{barticle}
\endbibitem

\bibitem[\protect\citeauthoryear{{Brooks} and
  {Warren}}{2012}]{2012ApJ...760L...5B}
\begin{barticle}
\bauthor{\bsnm{{Brooks}}, \binits{D.H.}},
\bauthor{\bsnm{{Warren}}, \binits{H.P.}}:
\byear{2012},
\batitle{{The Coronal Source of Extreme-ultraviolet Line Profile Asymmetries in
  Solar Active Region Outflows}}.
\bjtitle{\apjl}
\bvolume{760},
\bfpage{L5}.
\doiurl{10.1088/2041-8205/760/1/L5}.
\adsurl{2012ApJ...760L...5B}.
\end{barticle}
\endbibitem

\bibitem[\protect\citeauthoryear{{Bryans}, {Young}, and
  {Doschek}}{2010}]{2010ApJ...715.1012B}
\begin{barticle}
\bauthor{\bsnm{{Bryans}}, \binits{P.}},
\bauthor{\bsnm{{Young}}, \binits{P.R.}},
\bauthor{\bsnm{{Doschek}}, \binits{G.A.}}:
\byear{2010},
\batitle{{Multiple Component Outflows in an Active Region Observed with the EUV
  Imaging Spectrometer on Hinode}}.
\bjtitle{\apj}
\bvolume{715},
\bfpage{1012}.
\doiurl{10.1088/0004-637X/715/2/1012}.
\adsurl{2010ApJ...715.1012B}.
\end{barticle}
\endbibitem

\bibitem[\protect\citeauthoryear{{Brynildsen}, {Kjeldseth-Moe}, and
  {Maltby}}{1995}]{1995ApJ...455L..81B}
\begin{barticle}
\bauthor{\bsnm{{Brynildsen}}, \binits{N.}},
\bauthor{\bsnm{{Kjeldseth-Moe}}, \binits{O.}},
\bauthor{\bsnm{{Maltby}}, \binits{P.}}:
\byear{1995},
\batitle{{Quiet-Sun Connection between Intensity, Doppler Shift, and Line
  Broadening in Solar Ultraviolet Emission Lines}}.
\bjtitle{\apjl}
\bvolume{455},
\bfpage{L81}.
\doiurl{10.1086/309800}.
\adsurl{1995ApJ...455L..81B}.
\end{barticle}
\endbibitem

\bibitem[\protect\citeauthoryear{{Chae}, {Yun}, and
  {Poland}}{1998}]{1998ApJS..114..151C}
\begin{barticle}
\bauthor{\bsnm{{Chae}}, \binits{J.}},
\bauthor{\bsnm{{Yun}}, \binits{H.S.}},
\bauthor{\bsnm{{Poland}}, \binits{A.I.}}:
\byear{1998},
\batitle{{Temperature Dependence of Ultraviolet Line Average Doppler Shifts in
  the Quiet Sun}}.
\bjtitle{\apjs}
\bvolume{114},
\bfpage{151}.
\doiurl{10.1086/313064}.
\adsurl{1998ApJS..114..151C}.
\end{barticle}
\endbibitem

\bibitem[\protect\citeauthoryear{{Chandrasekhar}
  \textit{et~al.}}{1984}]{1984ApOpt..23..508C}
\begin{barticle}
\bauthor{\bsnm{{Chandrasekhar}}, \binits{T.}},
\bauthor{\bsnm{{Ashok}}, \binits{N.M.}},
\bauthor{\bsnm{{Desai}}, \binits{J.N.}},
\bauthor{\bsnm{{Pasachoff}}, \binits{J.M.}},
\bauthor{\bsnm{{Sivaraman}}, \binits{K.R.}}:
\byear{1984},
\batitle{{Fabry-Perot interferometric observations of the coronal red and green
  lines during the 1983 Indonesian eclipse}}.
\bjtitle{\appo}
\bvolume{23},
\bfpage{508}.
\doiurl{10.1364/AO.23.000508}.
\adsurl{1984ApOpt..23..508C}.
\end{barticle}
\endbibitem

\bibitem[\protect\citeauthoryear{{Chandrasekhar}
  \textit{et~al.}}{1991}]{1991SoPh..131...25C}
\begin{barticle}
\bauthor{\bsnm{{Chandrasekhar}}, \binits{T.}},
\bauthor{\bsnm{{Desai}}, \binits{J.N.}},
\bauthor{\bsnm{{Ashok}}, \binits{N.M.}},
\bauthor{\bsnm{{Pasachoff}}, \binits{J.M.}}:
\byear{1991},
\batitle{{Fabry-Perot line profiles in the 5303 A and 6374 A coronal lines
  obtained during the 1983 Indonesian eclipse}}.
\bjtitle{\solphys}
\bvolume{131},
\bfpage{25}.
\doiurl{10.1007/BF00151741}.
\adsurl{1991SoPh..131...25C}.
\end{barticle}
\endbibitem

\bibitem[\protect\citeauthoryear{{Dadashi}, {Teriaca}, and
  {Solanki}}{2011}]{2011A&A...534A..90D}
\begin{barticle}
\bauthor{\bsnm{{Dadashi}}, \binits{N.}},
\bauthor{\bsnm{{Teriaca}}, \binits{L.}},
\bauthor{\bsnm{{Solanki}}, \binits{S.K.}}:
\byear{2011},
\batitle{{The quiet Sun average Doppler shift of coronal lines up to 2 MK}}.
\bjtitle{\aap}
\bvolume{534},
\bfpage{A90}.
\doiurl{10.1051/0004-6361/201117234}.
\adsurl{2011A\%26A...534A..90D}.
\end{barticle}
\endbibitem

\bibitem[\protect\citeauthoryear{{De Pontieu}
  \textit{et~al.}}{2009}]{2009ApJ...701L...1D}
\begin{barticle}
\bauthor{\bsnm{{De Pontieu}}, \binits{B.}},
\bauthor{\bsnm{{McIntosh}}, \binits{S.W.}},
\bauthor{\bsnm{{Hansteen}}, \binits{V.H.}},
\bauthor{\bsnm{{Schrijver}}, \binits{C.J.}}:
\byear{2009},
\batitle{{Observing the Roots of Solar Coronal Heating{\mdash}in the
  Chromosphere}}.
\bjtitle{\apjl}
\bvolume{701},
\bfpage{L1}.
\doiurl{10.1088/0004-637X/701/1/L1}.
\adsurl{2009ApJ...701L...1D}.
\end{barticle}
\endbibitem

\bibitem[\protect\citeauthoryear{{Delone} and
  {Makarova}}{1969}]{1969SoPh....9..116D}
\begin{barticle}
\bauthor{\bsnm{{Delone}}, \binits{A.B.}},
\bauthor{\bsnm{{Makarova}}, \binits{E.A.}}:
\byear{1969},
\batitle{{Interferometric Investigation of the Red and Green Coronal Lines
  During the Total Solar Eclipse of May 30, 1965}}.
\bjtitle{\solphys}
\bvolume{9},
\bfpage{116}.
\doiurl{10.1007/BF00145733}.
\adsurl{1969SoPh....9..116D}.
\end{barticle}
\endbibitem

\bibitem[\protect\citeauthoryear{{Delone} and
  {Makarova}}{1975}]{1975SoPh...45..157D}
\begin{barticle}
\bauthor{\bsnm{{Delone}}, \binits{A.B.}},
\bauthor{\bsnm{{Makarova}}, \binits{E.A.}}:
\byear{1975},
\batitle{{Interferometric investigation of the line of sight velocities in the
  5303-A line during the eclipse of 11 September, 1968}}.
\bjtitle{\solphys}
\bvolume{45},
\bfpage{157}.
\doiurl{10.1007/BF00152228}.
\adsurl{1975SoPh...45..157D}.
\end{barticle}
\endbibitem

\bibitem[\protect\citeauthoryear{{Delone}, {Makarova}, and
  {Yakunina}}{1988}]{1988JApA....9..125D}
\begin{barticle}
\bauthor{\bsnm{{Delone}}, \binits{A.B.}},
\bauthor{\bsnm{{Makarova}}, \binits{E.A.}},
\bauthor{\bsnm{{Yakunina}}, \binits{G.V.}}:
\byear{1988},
\batitle{{Erratum: ``Evidence for moving features in the corona from emission
  line profiles observed during eclipses'' [J. Astrophys. Astron., Vol. 9, No.
  1, p. 41 - 47 (1988)].}}
\bjtitle{Journal of Astrophysics and Astronomy}
\bvolume{9},
\bfpage{125}.
\doiurl{10.1007/BF02715689}.
\adsurl{1988JApA....9..125D}.
\end{barticle}
\endbibitem

\bibitem[\protect\citeauthoryear{{McIntosh}
  \textit{et~al.}}{2012}]{2012ApJ...749...60M}
\begin{barticle}
\bauthor{\bsnm{{McIntosh}}, \binits{S.W.}},
\bauthor{\bsnm{{Tian}}, \binits{H.}},
\bauthor{\bsnm{{Sechler}}, \binits{M.}},
\bauthor{\bsnm{{De Pontieu}}, \binits{B.}}:
\byear{2012},
\batitle{{On the Doppler Velocity of Emission Line Profiles Formed in the
  ``Coronal Contraflow'' that Is the Chromosphere-Corona Mass Cycle}}.
\bjtitle{\apj}
\bvolume{749},
\bfpage{60}.
\doiurl{10.1088/0004-637X/749/1/60}.
\adsurl{2012ApJ...749...60M}.
\end{barticle}
\endbibitem

\bibitem[\protect\citeauthoryear{{Mierla}
  \textit{et~al.}}{2008}]{2008A&A...480..509M}
\begin{barticle}
\bauthor{\bsnm{{Mierla}}, \binits{M.}},
\bauthor{\bsnm{{Schwenn}}, \binits{R.}},
\bauthor{\bsnm{{Teriaca}}, \binits{L.}},
\bauthor{\bsnm{{Stenborg}}, \binits{G.}},
\bauthor{\bsnm{{Podlipnik}}, \binits{B.}}:
\byear{2008},
\batitle{{Analysis of the Fe X and Fe XIV line width in the solar corona using
  LASCO-C1 spectral data}}.
\bjtitle{\aap}
\bvolume{480},
\bfpage{509}.
\doiurl{10.1051/0004-6361:20078329}.
\adsurl{2008A\%26A...480..509M}.
\end{barticle}
\endbibitem

\bibitem[\protect\citeauthoryear{{Patsourakos} and
  {Klimchuk}}{2006}]{2006ApJ...647.1452P}
\begin{barticle}
\bauthor{\bsnm{{Patsourakos}}, \binits{S.}},
\bauthor{\bsnm{{Klimchuk}}, \binits{J.A.}}:
\byear{2006},
\batitle{{Nonthermal Spectral Line Broadening and the Nanoflare Model}}.
\bjtitle{\apj}
\bvolume{647},
\bfpage{1452}.
\doiurl{10.1086/505517}.
\adsurl{2006ApJ...647.1452P}.
\end{barticle}
\endbibitem

\bibitem[\protect\citeauthoryear{{Peter}}{2010}]{2010A&A...521A..51P}
\begin{barticle}
\bauthor{\bsnm{{Peter}}, \binits{H.}}:
\byear{2010},
\batitle{{Asymmetries of solar coronal extreme ultraviolet emission lines}}.
\bjtitle{\aap}
\bvolume{521},
\bfpage{A51}.
\doiurl{10.1051/0004-6361/201014433}.
\adsurl{2010A\%26A...521A..51P}.
\end{barticle}
\endbibitem

\bibitem[\protect\citeauthoryear{{Peter} and
  {Judge}}{1999}]{1999ApJ...522.1148P}
\begin{barticle}
\bauthor{\bsnm{{Peter}}, \binits{H.}},
\bauthor{\bsnm{{Judge}}, \binits{P.G.}}:
\byear{1999},
\batitle{{On the Doppler Shifts of Solar Ultraviolet Emission Lines}}.
\bjtitle{\apj}
\bvolume{522},
\bfpage{1148}.
\doiurl{10.1086/307672}.
\adsurl{1999ApJ...522.1148P}.
\end{barticle}
\endbibitem

\bibitem[\protect\citeauthoryear{{Prabhakar}, {Raju}, and
  {Chandrasekhar}}{2013}]{2013ASInC..10..137P}
\begin{bchapter}
\bauthor{\bsnm{{Prabhakar}}, \binits{M.}},
\bauthor{\bsnm{{Raju}}, \binits{K.P.}},
\bauthor{\bsnm{{Chandrasekhar}}, \binits{T.}}:
\byear{2013},
\bctitle{{Analysis of the solar coronal green line profiles from eclipse
  observations}}.
In: \bbtitle{Astronomical Society of India Conference Series},
\bsertitle{Astronomical Society of India Conference Series}
\bseriesno{10}.
\adsurl{2013ASInC..10..137P}.
\end{bchapter}
\endbibitem

\bibitem[\protect\citeauthoryear{{Prasad}, {Singh}, and
  {Banerjee}}{2013}]{2013SoPh..282..427P}
\begin{barticle}
\bauthor{\bsnm{{Prasad}}, \binits{S.K.}},
\bauthor{\bsnm{{Singh}}, \binits{J.}},
\bauthor{\bsnm{{Banerjee}}, \binits{D.}}:
\byear{2013},
\batitle{{Variation of Emission Line Width in Mid- and High-Latitude Corona}}.
\bjtitle{\solphys}
\bvolume{282},
\bfpage{427}.
\doiurl{10.1007/s11207-012-0160-0}.
\adsurl{2013SoPh..282..427P}.
\end{barticle}
\endbibitem

\bibitem[\protect\citeauthoryear{{Raju}}{1999}]{1999SoPh..185..311R}
\begin{barticle}
\bauthor{\bsnm{{Raju}}, \binits{K.P.}}:
\byear{1999},
\batitle{{The Effect of Mass Motions Inside the Coronal Loops on Emission Line
  Profiles}}.
\bjtitle{\solphys}
\bvolume{185},
\bfpage{311}.
\doiurl{10.1023/A:1005128428916}.
\adsurl{1999SoPh..185..311R}.
\end{barticle}
\endbibitem

\bibitem[\protect\citeauthoryear{{Raju}, {Chandrasekhar}, and
  {Ashok}}{2011}]{2011ApJ...736..164R}
\begin{barticle}
\bauthor{\bsnm{{Raju}}, \binits{K.P.}},
\bauthor{\bsnm{{Chandrasekhar}}, \binits{T.}},
\bauthor{\bsnm{{Ashok}}, \binits{N.M.}}:
\byear{2011},
\batitle{{Analysis of Coronal Green Line Profiles: Evidence of Excess
  Blueshifts}}.
\bjtitle{\apj}
\bvolume{736},
\bfpage{164}.
\doiurl{10.1088/0004-637X/736/2/164}.
\adsurl{2011ApJ...736..164R}.
\end{barticle}
\endbibitem

\bibitem[\protect\citeauthoryear{{Raju}
  \textit{et~al.}}{1993}]{1993MNRAS.263..789R}
\begin{barticle}
\bauthor{\bsnm{{Raju}}, \binits{K.P.}},
\bauthor{\bsnm{{Desai}}, \binits{J.N.}},
\bauthor{\bsnm{{Chandrasekhar}}, \binits{T.}},
\bauthor{\bsnm{{Ashok}}, \binits{N.M.}}:
\byear{1993},
\batitle{{Line-Of Velocities Observed in the Inner Solar Corona during the
  Total Solar Eclipses of 1980 and 1983}}.
\bjtitle{\mnras}
\bvolume{263},
\bfpage{789}.
\doiurl{10.1093/mnras/263.3.789}.
\adsurl{1993MNRAS.263..789R}.
\end{barticle}
\endbibitem

\bibitem[\protect\citeauthoryear{{Raouafi} and
  {Solanki}}{2004}]{2004A&A...427..725R}
\begin{barticle}
\bauthor{\bsnm{{Raouafi}}, \binits{N.-E.}},
\bauthor{\bsnm{{Solanki}}, \binits{S.K.}}:
\byear{2004},
\batitle{{Effect of the electron density stratification on off-limb O VI line
  profiles: How large is the velocity distribution anisotropy in the solar
  corona?}}
\bjtitle{\aap}
\bvolume{427},
\bfpage{725}.
\doiurl{10.1051/0004-6361:20041203}.
\adsurl{2004A\%26A...427..725R}.
\end{barticle}
\endbibitem

\bibitem[\protect\citeauthoryear{{Sakurai}}{2017}]{2017PJAB...93...87S}
\begin{barticle}
\bauthor{\bsnm{{Sakurai}}, \binits{T.}}:
\byear{2017},
\batitle{{Heating mechanisms of the solar corona}}.
\bjtitle{Proceeding of the Japan Academy, Series B}
\bvolume{93},
\bfpage{87}.
\doiurl{10.2183/pjab.93.006}.
\adsurl{2017PJAB...93...87S}.
\end{barticle}
\endbibitem

\bibitem[\protect\citeauthoryear{{Singh}
  \textit{et~al.}}{2006}]{2006JApA...27..115S}
\begin{barticle}
\bauthor{\bsnm{{Singh}}, \binits{J.}},
\bauthor{\bsnm{{Sakurai}}, \binits{T.}},
\bauthor{\bsnm{{Ichimoto}}, \binits{K.}},
\bauthor{\bsnm{{Muneer}}, \binits{S.}}:
\byear{2006},
\batitle{{Spectroscopic Studies of Solar Corona VI: Trend in Line-width
  Variation of Coronal Emission Lines with Height Independent of the Structure
  of Coronal Loops}}.
\bjtitle{Journal of Astrophysics and Astronomy}
\bvolume{27},
\bfpage{115}.
\doiurl{10.1007/BF02702514}.
\adsurl{2006JApA...27..115S}.
\end{barticle}
\endbibitem

\bibitem[\protect\citeauthoryear{{Teriaca}
  \textit{et~al.}}{1999}]{1999ESASP.446..645T}
\begin{bchapter}
\bauthor{\bsnm{{Teriaca}}, \binits{L.}},
\bauthor{\bsnm{{Banerjee}}, \binits{D.}},
\bauthor{\bsnm{{Doyle}}, \binits{J.G.}},
\bauthor{\bsnm{{Erd{\'e}ly}}, \binits{R.}}:
\byear{1999},
\bctitle{{SUMER Observations of Line Shifts in the Quiet Sun and in an Active
  Region}}.
In: \beditor{\bsnm{{Vial}}, \binits{J.-C.}},
\beditor{\bsnm{{Kaldeich-Sch{\"u}}}, \binits{B.}} (eds.)
\bbtitle{8th SOHO Workshop: Plasma Dynamics and Diagnostics in the Solar
  Transition Region and Corona},
\bsertitle{ESA Special Publication}
\bseriesno{446},
\bfpage{645}.
\adsurl{1999ESASP.446..645T}.
\end{bchapter}
\endbibitem

\end{thebibliography}

\end{article} 

\end{document}